\documentclass[%
 reprint,
nofootinbib,
 amsmath,amssymb,
 aps,
]{revtex4-2}

\usepackage{graphicx}
\usepackage{dcolumn}
\usepackage{bm}

\begin{document}

\preprint{APS/123-QED}

\title{Quantum Tunneling in SO(3,1)$\times$SO(4) }

\author{Robert Hipple}
\affiliation{%
Department of Electrical and Computer Engineering\\
Michigan State University\\
East Lansing, Michigan\\
}%

\date{\today}

\begin{abstract}
We supplement the Lorentz transformation $L(v)$ with a new ``Tunneling" transformation $T(v)$. Application of this new 
operation to elementary quantum mechanics offers a novel, intuitive insight into the nature of quantum tunneling; in particular, the so-called ``Klein Paradox" is discussed.

\end{abstract}

\pacs{Valid PACS appear here}
\maketitle


\section{\label{sec:level1}Introduction:\\Assumption of Spacetime Isotropy}
The Lorentz transformation admits eigenvectors \cite{Byron}. These eigenvectors correspond to definite directions in spacetime, which would seem to imply that there exist preferred directions in spacetime. For example, consider two laboratories, $A$ and $B$,  sufficiently separated in spacetime so that no signal traveling at the speed of light has yet been able to traverse the intervening space. If $A$ performs experiments to determine the direction in spacetime of the eigenvector of the Lorentz transformation (i.e., the speed of light), will $B$ measure the same direction in spacetime for his Lorentz transformation eigenvector?
\\

As a more specific example, consider a copy of $\mathbb{R}^2$ endowed with a cartesian coordinate system. Define a linear transformation 
\begin{equation}\nonumber
A:\mathbb{R}^2\rightarrow \mathbb{R}^2,
\end{equation}
and assume $A$ to have two distinct, real eigenvectors. There is an ambiguity in this definition due to the freedom in selection of coordinate system. For if 
\begin{equation}\nonumber
A{\bf x}=k{\bf x},
\end{equation}
and $R(\theta)$ represents a rotation about the angle $\theta$,
then
\begin{equation}\nonumber
\begin{split}
R(\theta)Ax=R(\theta)AR^{-1}(\theta)R(\theta)x=kR(\theta)x.
\end{split}
\end{equation}
Hence a rotation of the coordinate system results in similarly rotated eigenvectors.

The Lorentz transformation for motion in the $x$ direction with velocity $v$ can be  written in the form 
\begin{equation}\nonumber
L(v)=
\begin{bmatrix}
\gamma&0&0&-\gamma v\\
0&1&0&0\\
0&0&1&0\\
-\gamma\frac{v}{c^2}&0&0&\gamma
\end{bmatrix},
\end{equation}
where, as usual, 
\begin{equation}\nonumber
\gamma=\frac{1}{\sqrt{1-\frac{v^2}{c^2}}}.
\end{equation}
If we use the coordinate system $(x,y,z,ct)$ and set $\tanh{\eta}=v/c$, this transformation can be written as
\begin{equation}
L(\eta)=
\begin{bmatrix}
\cosh{\eta}&0&0&-\sinh{\eta}\\
0&1&0&0\\
0&0&1&0\\
-\sinh{\eta}&0&0&\cosh{\eta}
\end{bmatrix}.
\label{eq:lt_eta}
\end{equation}
Let us define a hypothetical ``spacetime rotation" by angle $\theta$ in the $(x,ct)$ plane:
\begin{equation}
R(\theta)=
\begin{bmatrix}
\cos{\theta}&0&0&\sin{\theta}\\
0&1&0&0\\
0&0&1&0\\
-\sin{\theta}&0&0&\cos{\theta}
\end{bmatrix}.
\label{eq:spacetimerot}
\end{equation}
which is different from a purely spatial rotation in that it involves the time coordinate.
Using Eq.~(\ref{eq:lt_eta}) as a guide, we define $\tan{\theta}=v/c$. This allows us to write Eq.~\ref{eq:spacetimerot} in the form
\begin{equation}\nonumber
T(v)=
\begin{bmatrix}
\gamma^+&0&0&\gamma^+v\\
0&1&0&0\\
0&0&1&0\\
-\gamma^+\frac{v}{c^2}&0&0&\gamma^+
\end{bmatrix},
\end{equation}
where
\begin{equation}\nonumber
\gamma^+=\frac{1}{\sqrt{1+\frac{v^2}{c^2}}}.
\end{equation}
%

The physical reality of a transformation such as $T(v)$ must be questioned. To pursue this inquiry, let $O_T$ be the frame of reference of an observer who was initially at rest, then subjected to a $T(v)$ transformation. The observer in the $O_T$ frame might perform experiments to determine the eigenvectors of a Lorentz transformation in his frame. That is, he would look for eigenvectors of   
\begin{equation}\nonumber
T(v)L(v')T(-v),
\end{equation}
and would measure them. These eigenvectors of the Lorentz transformation in the $O_T$ frame would constitute a ``speed of light" in that frame. Said another way,  the speed of light would be constant under Lorentz transformations \emph{in the $O_T$ frame}. Since the constancy of the speed of light in an inertial frame is a principle fundamental to Special Relativity, it is therefore seen that Special Relativity carries over unaltered into the ``spacetime-rotated" frame of $O_T$.

\section{\label{sec:level1}Existence of the hypothesized ``Spacetime Rotation" ${\bf T}(v)$}
If such a transformation $T(v)$ is possible, then there must be some means by which its effects can be detected. It will be shown that such a transformation does indeed exist, and that its effects are well known. 

To this end, introduce three observers: $O$, $O_L$ and $O_T$, sharing a common Cartesian coordinate system. Each of these observers are initially positioned at the origin. Suppose $O$ to be at rest, and let $O_L$ be subjected to a Lorentz transformation $L(v)$ with respect to $O$. Likewise, let $O_T$ be subjected to a transformation $T(v)$ with respect to $O$. The question before us is how might $O$ distinguish between $O_L$ and $O_T$?

\begin{figure}
\begin{center}
\includegraphics[width=.5\textwidth]{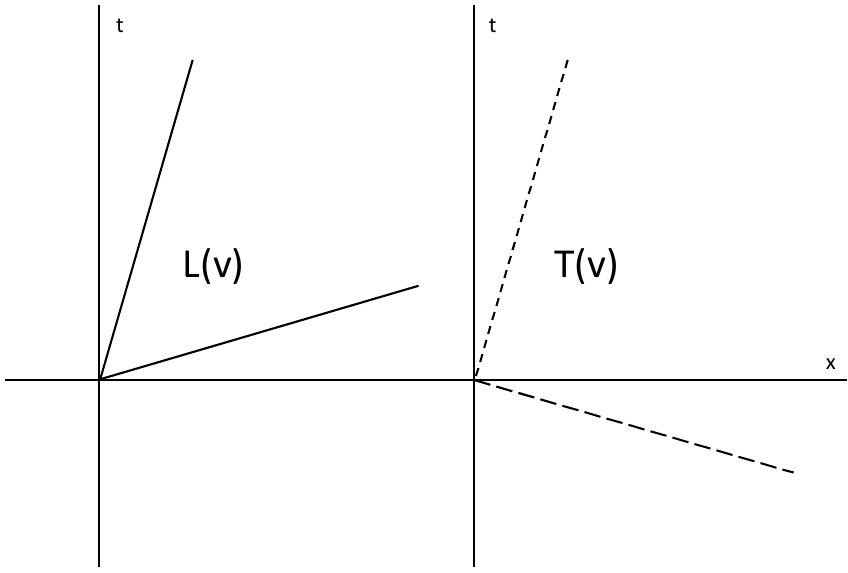}
\caption{ Comparison of $L(v)$ (solid lines) and $T(v)$ (dashed lines)}
\label{fig:phasevsgroup}
\end{center}
\end{figure}

It is sufficient to work in a single spatial dimension $x$ and time $t$. We have
\begin{equation}\nonumber
L(v)=\frac{1}{\sqrt{1-\frac{v^2}{c^2}}}
\begin{bmatrix}
1&-v\\
\frac{-v}{c^2}&1\\
\end{bmatrix},
\end{equation} 
and 
\begin{equation}\nonumber
T(v)=\frac{1}{\sqrt{1+\frac{v^2}{c^2}}}
\begin{bmatrix}
1&-v\\
\frac{v}{c^2}&1\\
\end{bmatrix}.
\end{equation} 
We note two differences between the $L(v)$ and $T(v)$ transformations;  the leading ``gamma" factors contain a sign difference, and the $v/c^2$ matrix elements have opposite signs. 
To describe the effects of these differences, we follow de Broglie \cite{deBroglie} and set up two simple wave functions for particles at rest - one for $O_L$: 
\begin{equation}
\Psi_L(x_L,t_L) =\cos\left(\frac{mc^2}{\hbar}t_L\right),
\label{eq:debroglieatrest}
\end{equation}
and another for $O_T$:
\begin{equation}\nonumber
\Psi_T(x_T,t_T) =\cos\left(\frac{mc^2}{\hbar}t_T\right).
\end{equation}
These are lowest order modes, independent of $x$.
We next apply the $L(v)$ and $T(v)$ transformations, respectively, to these wave functions to determine their appearance in the $O$ frame. We have
\begin{equation}\nonumber
t_L = \frac{t-vx/c^2}{\sqrt{1-v^2/c^2}},
\end{equation}
and
\begin{equation}\nonumber
t_T = \frac{t+vx/c^2}{\sqrt{1+v^2/c^2}},
\end{equation}
yielding
\begin{equation}\nonumber
\Psi_L = \cos\left(\frac{mc^2}{\hbar}\frac{t-vx/c^2}{\sqrt{1-v^2/c^2}}\right) = \cos(\omega_Lt-k_Lx),
\end{equation}
and
\begin{equation}\nonumber
\Psi_T = \cos\left(\frac{mc^2}{\hbar}\frac{t+vx/c^2}{\sqrt{1+v^2/c^2}}\right) = \cos(\omega_Tt+k_Tx).
\end{equation}
We have made the substitutions
\begin{equation}
k_L = \frac{mc^2}{\hbar}\frac{v/c^2}{\sqrt{1-v^2/c^2}},
\label{eq:KL}
\end{equation}
\begin{equation}
\omega_L = \frac{mc^2}{\hbar}\frac{1}{\sqrt{1-v^2/c^2}},
\label{eq:omegaL} 
\end{equation}
and
\begin{equation}
k_T = \frac{mc^2}{\hbar}\frac{v/c^2}{\sqrt{1+v^2/c^2}},
\label{eq:a}
\end{equation}
\begin{equation}
\omega_T = \frac{mc^2}{\hbar}\frac{1}{\sqrt{1+v^2/c^2}}.
\label{eq:b}
\end{equation}
We can discuss the phase and group velocities for $\Psi_L$ and $\Psi_T$ once we have determined their respective dispersion relations. Eliminating $v$ between (\ref{eq:KL}) and (\ref{eq:omegaL}), we find:
\begin{equation}\nonumber
\hbar^2\omega_L^2 = m^2c^4 + \hbar k_L^2c^2.
\end{equation}
Likewise for (\ref{eq:a}) and (\ref{eq:b}) we find:
\begin{equation}\nonumber
\hbar^2\omega_T^2 = m^2c^4 -\hbar k_T^2c^2.
\end{equation}
Next we calculate the phase and group velocities. For the Lorentz transformation case, $L(v)$, we derive the familiar relations
\begin{equation}
\text{Phase velocity: } \omega_L/k_L = c^2/v,
\label{eq:pvl}
\end{equation}
and
\begin{equation}
\text{Group velocity: } d\omega_L /dk_L = v.
\label{eq:gvl}
\end{equation}
In contrast, for the transformation $T(v)$ we derive
\begin{equation}
\text{Phase velocity: } \omega_T/k_T = c^2/v, 
\label{eq:pvt}
\end{equation}
and
\begin{equation}
\text{Group velocity: } d\omega_T/dk_T = -v.
\label{eq:gvt}
\end{equation}
This result highlights the essential difference between a Lorentz transformation $L(v)$ and a spacetime rotation $T(v)$; the phase velocities are identical, but the group velocities are oppositely directed. Equivalently, one can say that if two objects acquire the same velocity, one by virtue of applying $L(v)$ and the other by applying $T(v)$, they possess oppositely directed phase velocities (Fig.~\ref{fig:phasevsgroup}).
\\
The significance of this result can be made clear by examining the energy-momentum relation for the two cases;
\begin{equation}\nonumber
E_L^2 = m^2c^4 + p_L^2c^2,
\end{equation}
or, adding a potential $V$ \cite{Schiff}, 
\begin{equation}
(E_L-V)^2 = m^2c^4 + p_L^2c^2.
\label{eq:emlorentz}
\end{equation}
For small $p_L$ this becomes
\begin{equation}\nonumber
E_L-V = \frac{p_L^2}{2m},
\end{equation}
which is the nonrelativistic energy-momentum relation. Applying the standard prescription 
\begin{equation}
\begin{split}
E &\rightarrow i\hbar\frac{d}{dt}\\
p &\rightarrow -i\hbar\frac{d}{dx},
\end{split}
\label{eq:prescription}
\end{equation}
to convert this into a quantum mechanical statement we arrive at the nonrelativistic Shr\"odinger equation
\begin{equation}\nonumber
i\hbar\frac{\partial\Psi_L}{\partial t}=-\frac{\hbar^2}{2m}\frac{\partial^2\Psi_L}{dx^2}+V\Psi_L.
\end{equation}
Separating out the time dependence,
\begin{equation}
\frac{\hbar^2}{2m}\frac{d^2\psi_L(x)}{dx^2}+(E-V)\psi_L(x)=0,
\label{eq:lorentztimeindependentSE}
\end{equation}
we recognize the time independent Shr\"odinger equation. \\
Repeating the above procedure to derive the energy-momentum relation under the $T(v)$ transformation, we find:
\begin{equation}
(E_T-V)^2 = m^2c^4 - p_T^2c^2.
\label{eq:emrelationtunnel}
\end{equation}
Expanding this relation to second order about $p_T=0$ gives
\begin{equation}\nonumber
E_T-V = -\frac{p_T^2}{2m}.
\end{equation}
If we postulate that prescription (\ref{eq:prescription}) holds good for $T(v)$ transformed particles, we find the time-independent Schr\"odinger equation in this case takes the form
\begin{equation}
\frac{\hbar^2}{2m}\frac{d^2\psi_T(x)}{dx^2}+(V-E)\psi_T(x)=0.
\label{eq:tunneltimeindependentSE}
\end{equation}
Comparing (\ref{eq:lorentztimeindependentSE}) and (\ref{eq:tunneltimeindependentSE}), we clearly see the difference between a test particle whose rest frame has been ``Lorentz transformed" and another test particle whose rest frame has been transformed by $T(v)$. 

Transforming a particle by $T(v)$ is nothing less than the well-known phenomenon of \textbf{quantum tunneling}, seen here in a new light. This clearly demonstrates the physical reality of the newly proposed spacetime rotation, and suggests that we identify the new $T(v)$ transformation as the ``Tunneling" transformation since its effect is simply to transform a test particle from rest to a tunneling state.
\section{Commutativity of $\mathbf{L}$ and $\mathbf{T}$}
Typically one thinks of the state of a particle as ``tunneling" or ``not tunneling". But there is no reason why these transformations cannot be concatenated together. To explore this possibility, let us use the complex version of the Lorentz transformation \cite{Goldstein}, 
\begin{equation}
\mathbf{L}(\eta)=
\begin{bmatrix}
\cosh{\eta}&i\sinh{\eta}\\
i\sinh{\eta}&\cosh{\eta}
\end{bmatrix}.
\end{equation}
A Lorentz transform followed by a Tunneling transform is therefore
\begin{equation}
\begin{split}
    \mathbf{T(\eta)}\mathbf{L}(\theta)&=
    \begin{bmatrix}
    \cos{\theta}&\sin{\theta}\\
    -\sin{\theta}&\cos{\theta}
    \end{bmatrix}
    \begin{bmatrix}
    \cosh{\eta}&i\sinh{\eta}\\
     -i\sinh{\eta}&\cosh{\eta}
     \end{bmatrix}\\
    &=
     \begin{bmatrix}
     \cos{\theta}&\sin{\theta}\\
     -\sin{\theta}&\cos{\theta}
     \end{bmatrix}
     \begin{bmatrix}
     \cos{i\eta}&\sin{i\eta}\\
     -\sin{i\eta}&\cos{i\eta}
     \end{bmatrix}\\
    &=
     \begin{bmatrix}
     \cos{\theta+i\eta}&\sin{\theta+i\eta}\\
     -\sin{\theta+i\eta}&\cos{\theta+i\eta}
     \end{bmatrix}\\
     &=
     \begin{bmatrix}
     \cos{i\eta}&\sin{i\eta}\\
     -\sin{i\eta}&\cos{i\eta}
     \end{bmatrix}
     \begin{bmatrix}
     \cos{\theta}&\sin{\theta}\\
     -\sin{i\theta}&\cos{i\theta}
     \end{bmatrix}\\
     &=
     \begin{bmatrix}
     \cosh{\eta}&i\sinh{\eta}\\
     -i\sinh{\eta}&\cosh{\eta}
     \end{bmatrix}
     \begin{bmatrix}
     \cos{\theta}&\sin{\theta}\\
     -\sin{\theta}&\cos{\theta}
     \end{bmatrix}\\
     &=
     \mathbf{L}(\eta)\mathbf{T}(\theta)
\end{split}
\end{equation}
demonstrating that the  tunneling and Lorentz transformations combine unambiguously into a single well-defined ``generalized relativistic transformation". 

It is well-known that WKB problems can be treated by formally extending the domain of the wave function into complex variables. This allows one to circumvent singularities at turning points and caustics \cite{LL}. Here we see that wave functions ``extended into the complex plane" are hybrid states of tunneling and Lorentz transformed states. The velocity which is the parameter of tunneling and Lorentz transforms is seen to be a complex velocity $v=v_R+iv_I$, analogous to that from fluid mechanics. The ``real" velocity is the velocity associated with tunneling, while the ``imaginary" velocity is that associated with the Lorentz transformation.

\section{Tunneling in a Static, Homogeneous field.}
It will be shown that the Tunneling transformation arises from relativistic proper acceleration. Proper acceleration is defined as that which a test particle experiences within its rest frame \cite{Rindler}. The worldline of a test particle undergoing proper acceleration $\alpha$ is given by 
\begin{equation}\nonumber
x^2-c^2t^2=c^4/\alpha^2.
\end{equation}
As before, let the observer $O$ be at rest, and let $O_L$ be the Lorentz-transformed frame. We want $O_L$ to be uniformly accelerated by a constant value $\alpha$. We require the velocity of all points of $O_L$ to increase simultaneously. Since simultaneity is relative, we specify that this acceleration is with respect to the frame $O$; $v$ is then a function of $t$ alone,
\begin{equation}\nonumber
t'=\frac{t-v(t)x/c^2}{ \sqrt{1-\frac{v(t)^2}{c^2}}}=\frac{t-\alpha tx/c^2}{\sqrt{1-\alpha^2t^2/c^2}}.
\end{equation}
We want to calculate 
\begin{equation}\nonumber
\frac{\partial t'}{\partial t} = \frac{\partial}{\partial t}\left( \frac{t-v(t)x/c^2}{ \sqrt{1-\frac{v(t)^2}{c^2}}}\right)
=\frac{\partial}{\partial t}\left(\frac{t-\alpha tx/c^2}{\sqrt{1-\alpha^2t^2/c^2}}\right),
\end{equation} 
for fixed $x$, at time $t=0$. Further, we assume that $v=0$ initially. This gives
\begin{equation}\nonumber
\frac{\partial t'}{\partial t} = 1-\frac{\alpha x}{c^2}.
\end{equation} 
That is, $\partial t' / \partial t$ grows linearly with $x$, and  according to General Relativity, this is precisely the definition of a static homogeneous gravitational field \cite{Rohrlich} \cite{Moller} \cite{Einstein}.







%
Therefore, for constant acceleration in spacetime we can write
\begin{equation}\nonumber
t' = t-\alpha tx/c^2.
\end{equation}
This is nonlinear due to the $xt$ term, but if we expand $t'$ in the neighborhood of an event $(x_0,t_0)$ we can write
\begin{equation}\nonumber
dt'=\frac{\partial t'}{\partial x}dx+\frac{\partial t'}{\partial t}dt=-\frac{\alpha t_0}{c^2}dx+\left(1-\frac{\alpha x_0}{c^2}\right)dt.
\end{equation}
Repeating the earlier analysis of the de Broglie wave, we make the identifications
\begin{equation}
\begin{split}
\omega&=1-\frac{\alpha x_0 }{c^2},\\
k&=-\frac{\alpha t_0}{c^2},
\end{split}
\end{equation}
and calculate the phase velocity as
\begin{equation}\nonumber
v_\phi = \frac{1-\alpha x_0/c^2}{-\alpha t_0/c^2}.
\end{equation}
For group velocity we need $\omega(k)$. But there is no functional relationship between $\omega$ and $k$ unless there is a relationship between $x$ and $t$. We can create such a relation by specifying a path through spacetime and calculating the group velocity along that path\footnote{It has been pointed out \cite{Dykman} that since we are free to choose {\em any} values for $(x_0, t_0)$, one can find a group velocity of any value. Therefore we should only choose events which might actually describe the path of a particle.}. If we set $x(t) = vt$, for example, we find a group velocity of 
\begin{equation}\nonumber
v_g=\frac{d\omega}{dk}=v.
\end{equation} 
It remains to find the conditions by which the phase velocity $v_\phi$ and the group velocity $v_g$ have opposite directions, which is the indicator for the Tunneling transform. Since, for example, we are free to  choose $v_g>0$, we only require
\begin{equation}\nonumber
v_\phi = \frac{1-\alpha x_0/c^2}{-\alpha t_0/c^2} < 0,
\end{equation}
or 
\begin{equation}
\frac{c^2}{\alpha}<x_0.
\label{eq:tunnelcriterion}
\end{equation}
It is interesting to note that if we multiply (\ref{eq:tunnelcriterion}) by a parameter $m$, then this relation is equivalent to 
\begin{equation}\nonumber
mc^2 = m\alpha x.
\end{equation}
If $m$ is interpreted as the rest mass of a test particle, then $\alpha$ is identified as the acceleration of the static homogeneous field.
\section{Analysis of the Klein ``Paradox" }
The Klein ``Paradox" is not really a paradox at all \cite{Telegdi} \cite{Hansen}. It has long been recognized as the inadequacy of single particle quantum mechanics to describe what is actually a multiparticle process. A summary of the phenomenon is as follows. A test particle encounters a step potential, and the wave function splits into a reflected part which moves away from the step, and a transmitted part which tunnels into the step. The difficulty arises when the height of the potential step rises beyond a certain threshold, $V=2mc^2$, where $m$ is the rest mass of the test particle. It is instructive to analyze this process within the paradigm of the new Tunneling transform.
The initial work with the Klein Paradox was performed within the context of the Dirac equation \cite{Klein} \cite{Sauter}. However, it can also be analyzed in the context of the Klein-Gordon equation \cite{Winter}. 
We proceed as follows. From Schiff \cite{Schiff}, we start with the energy-momentum relation for $L(v)$:
\begin{equation}\nonumber
(E-V)^2=m^2c^4+p^2c^2.
\end{equation}
Assuming V to be a constant and making the substitutions (\ref{eq:prescription}) we derive the (one dimensional) Klein-Gordon equation for a test particle encountering a step potential: 
\begin{equation}\nonumber
\frac{d^2\phi}{dx^2} + \frac{(E-V)^2 - m^2c^4}{\hbar^2c^2}\phi=0.
\end{equation}
The solutions are oscillatory so long as 
\begin{equation}\nonumber
(E-V)^2 > m^2c^4.
\end{equation}
This means that the test particle must have enough energy after surmounting the potential $V$ to still possess a nonzero kinetic energy and also account for its rest energy. If 
\begin{equation}\nonumber
(E-V)^2=m^2c^4,
\end{equation}
then the test particle has no excess energy after accounting for the rest energy.  This is a turning point; the kinetic energy and momentum of the test particle at this point are $0$. 
If the potential is increased a bit more, then
\begin{equation}\nonumber
(E-V)^2<m^2c^4,
\end{equation}
and there is no excess energy with which to form momentum. To an observer at rest, the test particle is said to have entered a ``tunneling state" which is often interpreted as a state of ``imaginary momentum".
The Tunneling transform offers a more intuitive interpretation. Starting with the energy-momentum relation for $T(v)$ 
(\ref{eq:emrelationtunnel}):
\begin{equation}\nonumber
(E-V)^2=m^2c^4-p^2c^2,
\end{equation}
the Klein-Gordon equation for a test particle in a step potential now takes the following form:
\begin{equation}\nonumber
\frac{d^2\phi}{dx^2} + \frac{m^2c^4 - (E-V)^2 }{\hbar^2c^2}\phi=0.
\end{equation}
Refer to Fig. 2, which compares the energy-momentum relations for $L(v)$ and $T(v)$. We can interpret tunneling as the test particle converting its rest energy into a source of momentum. This can continue until all of the rest energy has been converted into momentum, i.e., when 
\begin{equation}\nonumber
(E-V)^2 =0,
\end{equation}
or
\begin{equation}
m^2c^4-p^2c^2=0. 
\end{equation}
\begin{figure}
\begin{center}
\includegraphics[width=0.5\textwidth]{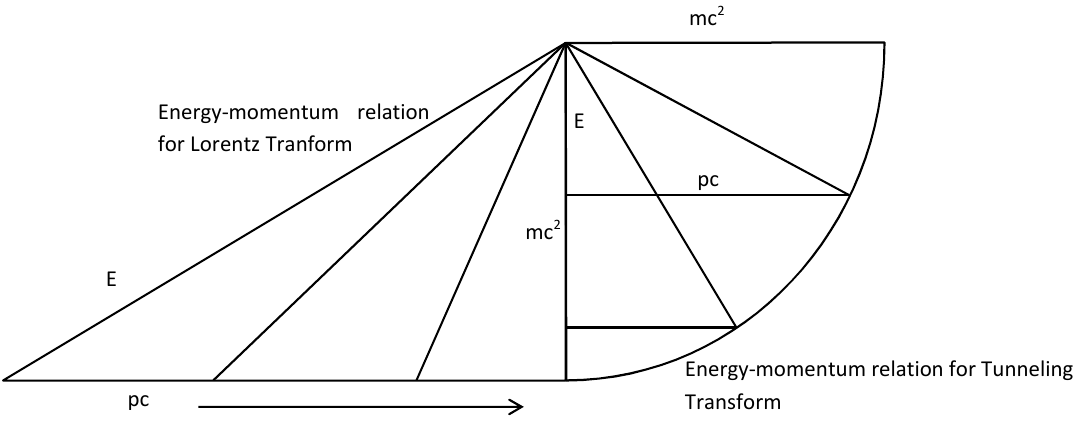}
\caption{Energy - momentum relations of $L(v)$ and $T(v)$}
\end{center}
\end{figure}
This is the largest value of momentum possible for this test particle. The entire kinetic energy as well as the rest energy of the particle has been converted into momentum; $pc=mc^2$.
Here we have only reached the point where $E=V$, or in the case of a test particle with no initial kinetic energy, just $V=mc^2$. At this point, the phase velocity is 0. The Tunneling transformation has rotated the frame of the test particle by $\pi/2$. Here the Tunneling transformation has the form 
\begin{equation}\nonumber
\begin{bmatrix}
\cos\frac{\pi}{2}&\sin\frac{\pi}{2}\\
-\sin\frac{\pi}{2}&\cos\frac{\pi}{2}\\
\end{bmatrix}
=
\begin{bmatrix}
0&1\\
-1&0\\
\end{bmatrix}.
\end{equation}
As we continue the Tunneling transform through larger angles, we move into the regime where we are on the negative branch of the energy \cite{Feynman}:
\begin{equation}\nonumber
E= -\sqrt{m^2c^4-p^2c^2},
\end{equation}
and $pc$ begins decreasing from its maximum value of $mc^2$. This is recognized as an antiparticle state; the Tunneling transform has continuously converted the test particle into an antiparticle. As $pc$ continues to decrease, it eventually reaches $pc=0$. It is at this point that we reach the threshold of the Klein paradox, an antiparticle at rest; i.e., 
\begin{equation}\nonumber
E-V = -\sqrt{m^2c^4},
\end{equation}
or
\begin{equation}
V > E + mc^2. 
\end{equation}
A final note about ``group velocity". The phase velocity of the wave function for a test particle at rest is infinite. This, of course, does not mean that any physical object is moving at an infinite rate of motion. It simply means that the set of events we are referring to appear to be simultaneous (the $x$-independent modulation of Eq.~(\ref{eq:debroglieatrest})) with respect to a particular observer. In the same way, what is a ``group velocity" to one observer can appear to be nothing more than a set of simultaneous events to another observer, so long as their respective frames are related by a Tunneling transformation. Therefore the apparent difficulties with a constant time rate gradient (e.g. ``infinitely large" group velocities) are not an issue when the frames are related by a Tunneling transformation. 

\section{Reversal of Transverse Magnetic Field}
The modification of electric and magnetic fields under a Lorentz transformation can best be described by the electromagnetic tensor \cite{wangsness},
\begin{equation}
F=
\begin{bmatrix}
0&B_z&-B_y&-i E_x/c\\
-B_z&0&B_x&-i E_y/c\\
B_y&-B_x&0&-i E_z/c\\
i E_x/c&i E_y/c&i E_z/c&0
\end{bmatrix}.
\label{eqn:emtensor}
\end{equation}
Let $L$ be the Lorentz transformation
\begin{equation}
    L=
    \begin{bmatrix}
    \gamma&0&0&i\beta\gamma\\
    0&1&0&0\\
    0&0&1&0\\
    -i\beta\gamma&0&0&\gamma
    \end{bmatrix}.
\end{equation}
Then the Lorentz transform of Eq.~\ref{eqn:emtensor} is given by
\begin{equation}
    F'=L^TFL.
\end{equation}
The individual components of the $E$, $B$ fields are transformed between the rest frame and the Lorentz frame as
\begin{equation}
\begin{split}
E'_x &= E_x,\\
E'_y &= \gamma(E_y-\beta cB_z),\\
E'_z &= \gamma(E_z+\beta cB_y),\\
B'_x &= B_x,\\
B'_y &= \gamma(B_y+\beta E_z/c),\\
B'_z &= \gamma(B_z-\beta E_y/c).
\end{split}
\label{eqn:emlorentz}
\end{equation}
Starting again with $F$, but this time applying the Tunneling transformation,
\begin{equation}\nonumber
T=
\begin{bmatrix}
\gamma&0&0&i\beta\gamma\\
0&1&0&0\\
0&0&1&0\\
i\beta\gamma&0&0&\gamma
\end{bmatrix},
\end{equation}
that is, by applying
\begin{equation}
    F'=T^TFT,
\end{equation}
to $F$,
results in the following relations between the $E$ and $B$ fields of the rest frame and of the tunneling frame:
\begin{equation}
\begin{split}
E'_x &= E_x\\
E'_y &= \gamma(E_y+\beta cB_z)\\
E'_z &= \gamma(E_z-\beta cB_y)\\
B'_x &= B_x\\
B'_y &= \gamma(B_y+\beta E_z/c)\\
B'_z &= \gamma(B_z-\beta E_y/c).
\end{split}
\label{eqn:emtunnel}
\end{equation}
Comparing equations (\ref{eqn:emlorentz}) and (\ref{eqn:emtunnel}) we find that most components of the electromagnetic field are transformed in the same manner by both transformations with the exception of $E'_x$ and $E'_y$. These components are transformed differently as if the transverse magnetic field of the rest frame had reversed sign. The conclusion is that a tunneling particle will respond to a transverse magnetic field in a manner opposite that of a Lorentz-transformed particle. This prediction could be experimentally verified.


\section{\label{sec:level1}Conclusion}
Originating from nothing more than symmetry considerations, we supplement the Lorentz transformation with a new ``Tunneling" transformation which is nothing more than a rotation in spacetime. Test particles subjected to this Tunneling transformation become tunneling particles; their essential property is that they experience a reversal in the direction of phase velocity with respect to group velocity. By extending our analysis to the complex plane, the Tunneling transform and the Lorentz transform are seen to commute. The combined transformation is a ``generalized relativistic transformation".

We have shown that the Tunneling transformation follows from the relativistic analysis of a static, homogeneous field. Finally, we apply the Tunneling transformation to the Klein paradox, and see that it is nothing more than the spacetime rotation of a tunneling particle into its antiparticle.

This work received no external funding.


\end{document}